\documentstyle[aps,pra,twocolumn]{revtex}

\begin{document}
\title{Relativity of motion in vacuum}
\author{Marc-Thierry Jaekel$^a$, Astrid Lambrecht$^b$
et Serge Reynaud$^b$}
\address{$a$ Laboratoire de Physique Th\'eorique de l'ENS
\thanks{Laboratoire du CNRS associ\'e \`a l'Ecole Normale
Su\-p\'e\-rieu\-re et \`a l'Uni\-ver\-si\-t\'e Paris-Sud},
24 rue Lhomond, F75231 Paris Cedex 05 \\
$b$ Laboratoire Kastler Brossel \thanks{Laboratoire de l'Ecole
Normale Su\-p\'e\-rieu\-re et de l'Uni\-ver\-si\-t\'e
Pierre et Marie Curie asso\-ci\'e au CNRS},
UPMC case 74, Jussieu, F75252 Paris Cedex 05\\}
\date{ Contribution for {\sc Vacuum}, eds E. Gunzig and S. Diner}

\maketitle

\bigskip

The principle of relativity is one of the main general laws of physics.
Since it applies in particular to motion in empty space
it is related to the symmetries of this space. Thoughts about this subject
within the framework of classical physics have led to the theory of
relativity \cite{EinsteinR}. The emergence of quantum theory has then
profoundly altered our conception of empty space by forcing us to
consider vacuum as the realm of quantum field fluctuations. The notion of
quantum vacuum and the existence of vacuum fluctuations unavoidably lead us
to reconsider the question of relativity of motion.
The present article is devoted to this aim with a main line which can
be formulated as follows:
{\it ``The principle of relativity of motion is directly
related to symmetries of quantum vacuum''}. Keeping close to this statement,
we discuss the controversial relation between vacuum
and motion. We introduce the question of relativity of motion
in its historical development before coming to the results
obtained more recently.

{\it Movement in space} cannot be defined otherwise than {\it movement in
vacuum}. This assertion was formulated by Leucippus and Democritus more
than 2000 years ago. In the context of that time, it has to be understood as
a logical necessity rather than a physical statement in the modern sense.
The existence of movement is a matter of evidence which forces us to
conceive a space in which movement takes place \cite{Russell}. The absence
of resistance to motion is a key condition in this respect and in fact
constitutes the main argument for the definition of the concept of vacuum.
At the same time vacuum is a positive physical reality which clearly
differs from nothingness. It is in particular the reference with respect
to which movement has to be identified.
These properties are certainly paradoxical at first sight.
This paradox has raised numerous discussions and
it has conserved its whole pertinence ever since the birth of modern
physics until today.

After a long eclipse dominated by Aristotelian concepts, the question of
relativity of motion was brought anew to the fore by Galileo \cite{Koyre}.
His central argument, ``{\it Movement is like nothing}'', not only signifies
that a motion with uniform velocity is indistinguishable from rest. It also
implies that a motion with uniform velocity has no observable effect when it
is composed with a second motion. This is the deep meaning of the famous
discussion of motion in a moving boat used by Galileo in the {\it Dialogo}
to emphasize the property of relativity \cite{Balibar}. Of course
this property is rigorously true only when the resistance of
air to motion can be ignored. Newton stressed this fact in the {\it %
Principia} in order to identify the empty space in which motion takes place
with
the vacuum which had just been the subject of the first experiments of
Torricelli, Pascal, von Guericke and Boyle. As we know today, the
Newtonian laws describing inertia or gravity already contain the
Galilean principle of relativity, as it was named by Einstein,
although Newton himself had argued for the absolute
character of his space reference. The theory of relativity
has freed space from this absolute character and one of the reasons
for that was precisely that it was contradictory with Galilean relativity.
After the advent of this theory, the relation between motion and space
can be made explicit as follows: the expression of
the laws of motion are symmetry properties of space or, equivalently,
the symmetries of empty space imply the laws of motion.

The formulation of the theory of relativity has largely been built upon the
symmetries of Maxwell's equations. Indeed the empty space of classical
physics is a reference for writing not only the laws of mechanics but also
the propagation of the electromagnetic field.
However, the development of classical physics was based on the
idealization that space can be thought as being absolutely empty.
This classical idealization could not be maintained, not even as a limiting
case,
when it was realized that space is always filled with freely propagating
radiation
fields after the birth of statistical mechanics and then of quantum mechanics.
Black body radiation is present at every non-zero temperature and it
undoubtedly produces real mechanical effects.
It exerts a pressure onto the reflecting boundaries of a surrounding cavity
and produces a friction force when reflecting surfaces are moving. Those two
effects are analogous to the pressure exerted by air molecules on the walls
of a container as was made clear by Einstein's theory of Brownian motion
\cite{EinsteinB}.
It is precisely for explaining the properties of black body radiation that
Planck introduced the first quantum law in 1900. At zero temperature
this law describes a region of space limited by a cavity and entirely
emptied out of radiation. In order to approach a
practical realization of vacuum, it is not sufficient to remove all matter
from this enclosure since one also has to lower the temperature down to
zero to eliminate thermal radiation.

Later on in 1912, Planck was led to modify his law in order
to improve its agreement with known results in the classical regime reached
at high temperatures. The modified law predicts that fluctuations remain
present at zero temperature, from which Nernst deduced in 1916 that
space is permanently filled with an electromagnetic field propagating with
the speed of light \cite{PlanckNernst}. The existence of these {\it %
zero-point fluctuations} or {\it vacuum fluctuations}, which correspond to
half of the energy of one photon per electromagnetic field mode, was
confirmed by the full quantum theory developed after 1925 \cite{DiracQM}.
Fluctuations thus appear as an inescapable consequence of Heisenberg's
inequalities. To sum up this lesson of quantum theory, one may say that
it establishes a necessary identity between the {\it potentiality}
of movement and the {\it presence} of movement. As soon as space allows
movement, which is of course an essential feature of space, then movement
exists and in particular subsists at zero temperature. In the same way, as
soon as space allows field propagation, which is equally a fundamental
property of space, then fields are present. Quantum physics forbids the
absence of movement as well as the absence of propagating fields. The ground
state of a mechanical oscillator is defined as the state where its
mechanical energy is minimal. In the same way, quantum vacuum is defined as
the field state where the energy of field fluctuations is minimal.

Then naturally the question arises whether the presence of fluctuations
in quantum vacuum is compatible with the principle of relativity of motion.
It is this discussion which we will sum up in a non-technical manner in this
article. More technical arguments as well as a list of references can be
found elsewhere \cite{RPP97}.

Some preliminary remarks should still be mentioned. Most of
the time the discussions devoted to Heisenberg's inequalities stress the
limits they impose on the precision of a measurement. Insisting on the
notions of {\it uncertainty} or even of {\it indeterminacy} however presents
the serious disadvantage that it fixes the ambition of physicists with
respect to a classical description of physical phenomena, although quantum
theory has been developed precisely in order to remove the deficiencies of
this description. The successes of quantum theoretical description of
natural phenomena plead for a quite renewed point of view. {\it Quantum
fluctuations} are intrinsic properties of physical quantities which display
their inherent quantum nature. Today's knowledge allows for a theoretical
understanding and also for an experimental study of these properties. In
certain
experiments one can even manipulate the quantum fluctuations in order to
reduce the noise they produce on a particular signal \cite{Squeezing}.

The fluctuations of the electromagnetic field which remain in the vacuum
state have well known observable effects \cite{CohenSciama}. An isolated
atom in vacuum interacts only with vacuum fluctuations and this interaction
is responsible for the {\it spontaneous emission} processes during which the
atom changes its internal state by falling from a higher energy level to a
lower one. When fallen in the ground state, the state of lowest energy where
it can no longer emit photons, the atom is still coupled to vacuum
fluctuations and this coupling results in measurable effects like the {\it %
Lamb shift} of the atomic absorption frequencies. Two atoms in
vacuum are coupled to the field fluctuations which
produce an attractive force between them, the so-called {\it Van der Waals
force}. This force plays a very important role in physical and chemical
processes and its quantum theoretical interpretation has been studied since
the first years of quantum theory. While studying this problem, Casimir
discovered in 1948 that there exists also a force between two mirrors placed
in quantum vacuum \cite{Casimir}. The vacuum fluctuations are modified by
the cavity formed by the mirrors and their energy depends on their
relative distance. Hence vacuum exerts a force which
mutually attracts the mirrors to each other. This {\it Casimir force,} as it
was called later on, depends only on the distance and on two
fundamental constants, the speed of light $c$ and the Planck constant $\hbar $.
This is a remarkably universal feature in particular because the Casimir
force is independent of the electronic charge in contrast to the Van der
Waals forces. We have already noticed that any pragmatic definition of
vacuum necessarily involves a region of space limited by some enclosure. The
Casimir effect signifies in fact that the energy of vacuum depends on the
configuration of this cavity from which it follows that its boundaries
experience forces arising from radiation pressure of vacuum fluctuations.
Although the Casimir force is relatively small it has been observed in several
experiments \cite{ManipCas}.

Vacuum fluctuations thus play a central role in the modern
description of the structure of matter. They explain numerous novel
effects which have been observed without any ambiguity. Their existence
is directly associated with a fundamental property of
quantum physics, namely the representation of any physical quantity by an
observable defined with the help of non-commuting mathematical objects.
Their status nevertheless continues to raise intricate questions. One
reason for this situation is the obvious fact that their representations
are often incompatible with the intuitions inherited from classical
physics. More fundamental reasons are related to the serious difficulties
which have remained unsolved for a long time and, for some of them, still
remain unsolved at the interface between the physics of vacuum fluctuations
and the laws of mechanics or gravitation.

The archetype of these difficulties is the relation of vacuum
energy to gravitation. The total vacuum energy, that is to say the energy
summed over all field modes in their vacuum state, takes an infinite value.
This implies that vacuum energy does not contribute in a standard way to
gravitation since the universe would have a very different appearance
otherwise. One simple way to deal with this problem is to set the vacuum
energy equal to zero and, therefore, to use it as a reference for all other
energies \cite{Enz}. In contrast to a widely spread opinion this
prescription does not allow to ignore the gravitational effects of vacuum.
Indeed vacuum energy is modified in a space curved by the gravitational
field. Furthermore, even if the mean value of the vacuum energy does not
contribute to gravitation its variations necessarily contribute to it. Yet,
the spatial distribution of vacuum energy changes continuously as it is
composed of field fluctuations which propagate with the speed of light. The
energy density of vacuum shows fluctuations which manifest themselves as
fluctuations of space-time curvature. It is possible to discuss some of
these questions in analogy with the standard formalism of quantum field
theory (a discussion and references can be found in \cite{AdP95}). However a
final answer to these questions will not be reached until a satisfactory
quantum description of gravitation is available.

As already stated in the introduction, from the mere point of view of logics
movement in space must be understood as movement in vacuum, so that the
presence of field fluctuations in quantum vacuum forces us to reconsider the
notion of motion itself. This question has been discussed at length in
connection with attempts to obtain a consistent description of motion for
elementary quantum objects like the electron \cite{Dirac38}. It has been
learned from classical electrodynamics that the expression of the force
acting on a moving charged particle contains a contribution, known as the
Abraham-Lorentz force, describing the reaction to motion entailed by the
electromagnetic field emitted by the particle. The modern quantum theory
tells us that the radiation reaction force is directly related to the
vacuum field fluctuations through the fluctuation-dissipation
relations, which are the quantum generalizations of the classical results of
Brownian motion theory \cite{Callen}. The occurence of a dissipative force
induced by motion of the electron and its direct association with vacuum
fluctuations forbid to obtain any consistent description of
atoms within the classical framework. In the quantum theory, this crisis was
only solved at a very high price since any mechanical description of
movement was abandoned for elementary quantum objects \cite{Heisenberg}.
Before discussing this point in more detail in the next
paragraphs, let us notice that this was not the end of the story. Vacuum
field fluctuations were also shown to modify the inertial mass of a
point-like scatterer and, furthermore, to lead to an infinite mass
correction. A renormalization prescription was designed, which states that
the {\it real} mass of the particle is finite, as the result of adding the
infinite positive correction to a {\it bare} mass which is itself infinitely
negative. However, other difficulties emerge as outcomes of this approach.
In particular, the mechanical response of the particle to an applied force
shows instability and violates causality \cite{Rohrlich}. In fact, the
renormalization procedure used to keep the particle mass finite in spite of
an infinite correction is incompatible with a causal mechanical description
\cite{Dekker}.

These difficulties have led most of the theoretical physicists to adopt a
pragmatic point of view which has proved itself useful in the
microscopic domain. Any mechanical description is then given up in this
domain where physical descriptions are instead built on quantum field
theory. Taken seriously, this approach forces physicists to renounce to
general principles of mechanics at the elementary level of microscopic
physics whereas the mechanical description of nature has a validity
restricted to the macroscopic domain \cite{Rosenfeld}. This {\it a priori}
separation between microscopic and macroscopic domains has now lost the
pragmatical pertinence it had when the argument was formulated, due to the
progress towards highly sensitive measurements approaching the quantum level
of sensitivity for macroscopic objects \cite{BKBO}. From a more fundamental
point of view, this solution cannot be satisfactory since it rejects the
mechanical questions which concern in particular inertia and gravitation
outside the framework of quantum theory. Although the problems of inertia
and gravitation differ, they are directly related to each other. In the
domain of gravitation also, vacuum field fluctuations modify the laws of
motion by the introduction of dissipative reaction terms which
lead to stability problems \cite{Fulling}. Both cases point at
difficulties generated by the mixture of classical and quantum theoretical
descriptions, namely Einstein or Newton equations on one hand and quantum
field theory on the other hand. It is known more generally that any
description built on a too crude mixing between classical and quantum
formalisms is necessarily plagued with inconsistencies \cite{deWitt62}.

The general problem of quantum field theory with moving boundaries has
emerged as an outcome of the thoughts about possible approaches to this
problem \cite{deWitt75}. This subject may be considered as an extension of
the study of Casimir force, precisely of vacuum radiation pressure on
reflecting boundaries, to the case of moving boundaries. At the same time, it
is directly related to the questions raised by quantum descriptions of
motion, inertia or gravitation. As it will become apparent in the following,
this problem has been given formulations well adapted to the present
theoretical framework of quantum physics and has led to satisfactory
solutions of some of the general difficulties mentioned above.

Let us first consider the simple case of a mirror moving with a uniform
velocity.
Clearly there exist two very different situations depending on whether
motion takes place in vacuum or in a thermal field. The principle of
relativity of motion applies to the first situation but not to the second
one. Should vacuum exert a force on a mirror with a uniform
velocity, the reaction of vacuum would distinguish between inertial motion
and rest. As expected, quantum theory predicts the friction
force to vanish in this case so that the principle of relativity
of uniform motion is valid in quantum vacuum. Besides, quantum theory gives an
interesting interpretation of this property according to which vacuum
fluctuations appear exactly identical to an inertial observer
and to an observer at rest. The invariance of vacuum
under Lorentz transformations is an essential condition for the principle
of relativity of motion and it establishes a precise relation between this
principle and a symmetry of vacuum.

We now come to the general case of arbitrary motion in vacuum. In 1976,
Fulling and Davies showed that a perfectly reflecting mirror moving in
vacuum emits radiation as soon as the mirror has a non uniform acceleration
\cite{FullingDavies}. Meanwhile, the motion tends to be reduced to a uniform
acceleration by the reaction of vacuum. The
change of mechanical energy associated with damping is dissipated in form of
emitted radiation and the reaction force follows
from the associated momentum exchange. This {\it motional force}
is directly related to fluctuations of the vacuum radiation pressure
as they may be evaluated for a mirror at rest, in full
consistency with the quantum fluctuation-dissipation relations
\cite{QO92,Barton}.
When the equations of motion in vacuum are modified in order to take the
motional force into account, it turns out that mirrors possess causal
mechanical response functions and hence stable motions in vacuum. It has to
be noted that the expression of the reaction force coincides with the
Abraham-Lorentz derivative in the limit of a perfectly reflecting mirror
which thus raises the same problems as in the case of the electron. This
difficulty is solved by a careful description of the reflection
properties of the mirror, accounting in particular for the fact that any
real mirror is certainly transparent at high frequencies.
The motion of a real mirror in vacuum
is described in a consistent manner once these points are
satisfactorily dealt with \cite{PLA92}. Furthermore, a full treatment of the
coupling between the mirror's motion and the field scattering shows that the
vacuum fluctuations of the quantum field are eventually transmitted to the
position of the mirror, even if the latter was originally considered as
classical. In any realistic situation, the mechanical effect of vacuum
upon the mirror may be considered as vanishingly small and the mirror is
found in this limit to obey the standard Schr\"{o}dinger equation \cite{JP93}.
These results prove that the objections against the
possibility of giving a consistent quantum representation of motion in vacuum
can be bypassed. They also show that the solution involves subtle correlations
between the descriptions of motion in vacuum and of vacuum itself. For
instance, the fluctuations associated with motion in vacuum can in no way be
treated as uncorrelated with the fields fluctuations of vacuum in which motion
takes place.

The existence of motional effects also questions the
principle of relativity of motion applied to arbitrary motions in vacuum.
As the reaction of vacuum vanishes for uniform
velocities it does not raise any problem to the theory of special relativity.
In contrast a non-uniformly accelerated motion produces observable effects,
namely the resistance of vacuum against motion and the emission of radiation
by the moving mirror into vacuum. Quantum theory thus allows us to sketch
the outline of a framework where the questions raised in the introduction
find consistent answers. Space is not empty since vacuum fluctuations
are always present. These fluctuations effectively represent a reference for
the definition of motion because they give rise to real dissipative effects
in the general case of an arbitrary motion. At the same time, they do not
damp uniform motions since quantum vacuum obeys Lorentz invariance.
The physical properties of quantum vacuum are thus consistent with
logical requirements of ancient atomists as well as with the Galilean
principle of relativity of motion.

In this context, it would be extremely interesting to obtain experimental
evidence for motional dissipative effects. Although these effects are
extremely small for any motion which could be achieved in practice for a
single mirror, an experimental observation could turn out to be conceivable
with a cavity, instead of a single mirror, oscillating in vacuum. Due to a
resonance enhancement of the emission of motional radiation the experimental
figures become indeed much more promising in this case \cite{PRL96}.

The radiation reaction force calculated by Fulling and Davies for a single
mirror in vacuum is proportional to the Abraham-Lorentz derivative
associated with its motion. Hence, it vanishes exactly for a
uniformly accelerated motion. One question then arises, which naturally
generalizes the one already met when considering uniform motion.
How do vacuum fluctuations appear for an observer with uniform
acceleration? The absence of reaction of vacuum against
uniformly accelerated motion would suit well the invariance of vacuum
fluctuations under a group of transformations corresponding to observers
with uniform acceleration as well as uniform velocity. This question has
raised and still raises numerous controversial discussions between
physicists \cite{Paradox}. To understand this controversy it is necessary to
describe what is often called the {\it Unruh effect} which predicts that
vacuum fluctuations should appear as thermal fluctuations in a uniformly
accelerated reference frame \cite{Unruh}. The effective temperature, which
is proportional to Planck constant and to acceleration, is analogous to the
Hawking temperature of the field radiated by the surface of a black hole
\cite{Hawking}. This analogy plays an important role in the arguments
pleading in favor of the physical reality of the Unruh effect but it can be
questioned as both situations are clearly different within the framework of
general relativity. The curvature of space-time is not involved in a change
of reference frame whereas it plays a central role in the description of a
black hole.

Concerning the problem of relativity of motion, the Unruh effect shows
serious drawbacks. If vacuum is really different for accelerated and
inertial observers, this certainly makes it difficult, if not impossible,
to describe inertia in a consistent quantum formalism. In particular
difficulties should arise when analysing situations which involve
composed motions. For example a motion with uniform velocity in an
accelerated frame should give rise to a vacuum reaction like uniform
velocity in an ordinary thermal field whereas a uniformly accelerated motion
in a frame obtained from a Lorentz transformation would be free from
dissipation. Contradictions are also met concerning energy conservation for
an accelerated detector. Furthermore, the interpretation of detection
processes appear to depend on the reference frame. As far as this point is
concerned different results have been obtained depending on the model used
for the detector \cite{Paradox}. For those models involving detectors which
click when accelerated in vacuum it can even happen that the detection
process is found to be related now to absorption then to emission of a
photon. More generally it is impossible to preserve the usual understanding
of the principle of relativity because the notions of vacuum or photon
number are not the same for different observers \cite{Ginzburg}.

These difficulties have their origin in a too rapid identification of
uniformly accelerated reference frames with their Rindler representation.
This particular representation of accelerated frames favors a criterion of
rigidity in the transformation of solid bodies. A consequence of this choice
is that Maxwell equations are modified in the change of reference frame.
The Unruh effect essentially tells us that the definition of vacuum is also
altered in a Rindler transformation and that the accelerated vacuum obtained
after the transformation appears as a thermal field.
But the Rindler representation of accelerated
reference frames is not the only possible one. In fact an infinite number of
representations exist when one only imposes the condition that a given point
with uniform acceleration is brought to rest. Furthermore, the theory of
general relativity does not provide any manner to privilegiate one
particular representation of accelerated frames since it convincingly argues
that physical results cannot depend on the choice of a particular map of
space-time.

In quantum theory in contrast, motion must be understood as taking
place in a space filled with vacuum fluctuations.
This feature implies to attach a particular importance to those transformations
to
accelerated frames which leave vacuum invariant. This leads to favor conformal
coordinate transformations which have been known for a long time to leave
Maxwell equations invariant \cite{BC}. These tranformations represent a
natural extension of Lorentz transformations which also include uniformly
accelerated frames \cite{FRW}. An essential property of conformal
transformations is to preserve the definition of vacuum fluctuations
\cite{QSO95} and, more generally, the definition of particle number
\cite{BJP95}. Hence there is no more difference between the viewpoints of
accelerated and of inertial observers as far as their perception of vacuum
is concerned, provided that accelerated observers are defined through
conformal transformations. The fact that vacuum exerts no
force on a uniformly accelerated object, already discussed for
electrons and for mirrors, is in fact a direct consequence of this symmetry
property of quantum vacuum. In this sense, the principle of relativity which
was known for uniform velocities now has a domain of application
extended to uniform accelerations.

In the conformal representation of accelerated frames, the Unruh effect
disappears as well as the paradoxes it creates. A deep reason for the
difficulties mentioned above is that Rindler
transformations do not form a group. In fact they do not even compose well with
Poincar\'{e} transformations. But the discussion of the principle of
relativity is mainly based on group properties obeyed by the
composition of motions and these arguments do not hold for Rindler
transformations. The situation is much more favorable for
conformal coordinate transformations which form a group including the
Poincar\'{e} transformations besides the conformal changes to accelerated
frames.
In particular, the composition of a uniformly accelerated motion with an
inertial motion results in another uniformly accelerated motion. Associated
with
preservation of Maxwell equations and of electromagnetic vacuum, these group
properties amply justify the use of conformal transformations
to represent uniformly accelerated reference frames.
The conformal representation is the natural choice
when relativistic properties are interpreted as invariance properties
rather than mere form-invariance relations \cite{Norton}.

Furthermore, a clear understanding of the notion of quantum
particles requires that the notion of vacuum has been understood
before. This point has been thoroughly discussed in the present paper
for what concerns the principle of relativity of motion.
It must equally be taken into account when considering the extension
of the equivalence principle to the quantum domain.
The thoughts presented here plead for
bringing out the symmetries of vacuum, the quantum version of empty
space, in the forefront of primary questions.
As it has been shown here, this should ensure compatibility of
the conception of motion with the principle of relativity.
It has also been shown elsewhere that this approach
fits perfectly well the description of localization in
space-time built upon symmetries of quantum fields.
Precisely, conformal symmetry allows to define quantum
observables associated with positions in space-time and
to obtain their transformations under Lorentz transformations
and transformations to accelerated frames \cite{JRloc}.

It has often been noted that the serious problems arising in the
mechanical description of electrons is directly associated with the fact
that they are treated as point-like structures although such a treatment
certainly contradicts their quantum nature.
More generally, the classical conception of space as a set of points
upon which classical relativity is built with the help of differential
geometry is challenged by the quantum nature of physical observables.
The design of new conceptions of motion and localization in space
could therefore reveal itself a preliminary step to the
solution of yet unsolved problems lying at the interface
between quantum theory and gravitation.


\bigskip

\begin{references}
\bibitem{EinsteinR}  Einstein A. 1946 {\it The Meaning of Relativity}
(Princeton University Press)

\bibitem{Russell}  Russell B. 1961 {\it History of Western Philosophy}
(George Allen \& Unwin; reprinted by Routledge 1991)

\bibitem{Koyre}  Koyr\'{e} A. 1957 {\it From the Closed World to the Open
Universe} (JohnHopkins Press)

\bibitem{Balibar}  Balibar F. 1984 {\it Galil\'{e}e, Newton lus par Einstein
Espace et relativit\'{e}} (Presses Universitaires de France); 1992 {\it %
Einstein 1905 De l'\'{e}ther aux quanta} (Presses Universitaires de France)

\bibitem{EinsteinB}  Einstein A. 1905 {\it Ann. Physik} {\bf 17} 549; 1909
{\it Phys. Z.} {\bf 10} 185; 1917 {\it Phys. Z.} {\bf 18} 121

\bibitem{PlanckNernst}  Planck M. 1900 {\it Verh. Deutsch. Phys. Ges.} {\bf %
2 } 237; 1911 {\it ibid.} {\bf 13} 138; Nernst W. 1916 {\it ibid.} {\bf 18}
83

\bibitem{DiracQM}  Dirac P. A. M. 1958 {\it The Principles of Quantum
Mechanics} (Oxford University Press)

\bibitem{RPP97}  Jaekel M. T. and Reynaud S. 1997 {\it Rep. Progr. Phys.}
{\bf 60} 863

\bibitem{Squeezing} Reynaud S., Heidmann A., Giacobino E. and Fabre C.
1992 in {\it Progress in Optics XXX} ed. E. Wolf (North Holland) 1;
Giacobino E. and Fabre C. (eds) 1992
{\it Quantum Noise Reduction in Optical Systems}
{\it Applied Physics} {\bf B 55} 189-303 (Special Issue);
Reynaud S., Giacobino E. and Zinn-Justin J. (eds) 1997
{\it Quantum Fluctuations, Proceedings of Les Houches Summer School LXIII}
(Elsevier)

\bibitem{CohenSciama}  Cohen-Tannoudji C., Dupont-Roc J. and Grynberg G.
1988 {\it Processus d'interaction entre photons et atomes}
(InterEditions) [1992 {\it Atom-Photon Interactions} (Wiley)];
Sciama D. W. 1991 in {\it The Philosophy of Vacuum} eds S.
Saunders and H. R. Brown (Clarendon)

\bibitem{Casimir}  Casimir H. B. G. 1948 {\it Proc. K. Ned. Akad. Wet.} {\bf %
51} 793

\bibitem{ManipCas}  Deriagin B. V. and Abrikosova I. I. 1957 {\it Jh. Eksp.
Teor. Fis.} {\bf 30} 993 [{\it Soviet Physics JETP} {\bf 3} 819]; Spaarnay
M. J. 1958 {\it Physica} {\bf XXIV}, 751; Tabor D. and Winterton R. H. S.
1968 {\it Nature} {\bf 219}, 1120; Black W., De Jong J. G. V., Overbeek J.
Th. G. and Sparnaay M. J.1968 {\it Trans. Faraday Soc.} {\bf 56} 1597;
Sabisky E. S. and Anderson C. H. 1973 {\it Phys. Rev.} {\bf A7} 790;
Lamoreaux S. K. 1997 {\it Phys. Rev. Lett.} {\bf 78}, 5

\bibitem{Enz}  Enz C. P. 1974 in {\it Physical Reality and Mathematical
Description} eds C. P. Enz and J. Mehra (Reidel)

\bibitem{AdP95}  Jaekel M. T. and Reynaud S. 1995 {\it Ann. Physik} {\bf 4}
68

\bibitem{Dirac38}  Dirac P. A. M. 1938 {\it Proc. Roy. Soc. London}
{\bf CLXVII A} 148

\bibitem{Callen}  Callen H. B. and Welton T. A. 1951 {\it Phys. Rev.} {\bf 83%
} 34; Kubo R. 1966 {\it Rep. Prog. Phys.} {\bf 29} 255

\bibitem{Heisenberg}  Heisenberg W. 1930 {\it The Physical Principles of
the Quantum Theory} (Dover)

\bibitem{Rohrlich}  Rohrlich F. 1965 {\it Classical Charged Particles}
(Addison Wesley)

\bibitem{Dekker}  Dekker H. 1985 {\it Phys. Lett.} {\bf A 107} 255; 1985
{\it Physica} {\bf 133 A} 1

\bibitem{Rosenfeld}  Rosenfeld L. 1963 {\it Nucl. Phys.} {\bf 40} 353

\bibitem{BKBO}  Braginsky V. B. and Khalili F. Ya. 1992 {\it Quantum
Measurement} (Cambridge University Press); Bocko M. F. and Onofrio R. 1996
{\it Rev. Mod. Phys.} {\bf 68} 755

\bibitem{Fulling}  Fulling S. A. 1989 {\it Aspects of Quantum Field Theory
in Curved Spacetime} (Cambridge University Press)

\bibitem{deWitt62}  de Witt B. S. 1962 {\it J. Math. Phys.} {\bf 3} 619

\bibitem{deWitt75}  de Witt B. S. 1975 {\it Phys. Rep.} {\bf 19} 295

\bibitem{FullingDavies}  Fulling S. A. and Davies P. C. W. 1976 {\it Proc.
R. Soc.} {\bf A 348} 393

\bibitem{QO92}  Jaekel M. T. and Reynaud S. 1992 {\it Quantum Opt.} {\bf 4}
39

\bibitem{Barton}   Barton G.
1994 {\it Cavity Quantum Electrodynamics (Supplement: Advances in
Atomic, Molecular and Optical Physics)} ed. P. Berman (Academic Press)

\bibitem{PLA92}  Jaekel M. T. and Reynaud S. 1992 {\it Phys. Lett.} {\bf A
167} 227

\bibitem{JP93}  Jaekel M. T. and Reynaud S. 1993 {\it J. Physique }{\bf I 3}
1

\bibitem{PRL96}  Lambrecht A., Jaekel M. T. and Reynaud S. 1996 {\it Phys.
Rev. Lett.} {\bf 77} 615

\bibitem{Paradox}  Unruh W. G. and Wald R. M. 1984 {\it Phys. Rev.} {\bf D
29 } 1047; Unruh W. G. 1992 {\it Phys. Rev.} {\bf D 46} 3271; Grove P. G.
1986 {\it Class. Quantum Grav.} {\bf 3} 793; 1988 {\it Class. Quantum Grav.}
{\bf 5} 1381; Raine D. J., Sciama D. W. and Grove P. G. 1991 {\it Proc. R.
Soc. London} {\bf A 435} 205

\bibitem{Unruh}  Davies P. C. W. 1975 {\it J. Physics} {\bf A 8} 609; Unruh
W. G. 1976 {\it Phys. Rev.} {\bf D 14} 870; Birrell N. D. and Davies P. C.
W. 1982 {\it Quantum Fields in Curved Space} (Cambridge University Press)

\bibitem{Hawking}  Hawking S. W. 1975 {\it Comm. Math. Phys.} {\bf 43} 199

\bibitem{Ginzburg}  Ginzburg V. L. and Frolov V. P. 1987 {\it Usp. Fiz. Nauk}
{\bf 153} 633 [{\it Sov. Phys. Usp.} {\bf 30} 1073]

\bibitem{BC}  Bateman H. 1909 {\it Proc. London Math. Soc.} {\bf 7} 70;
{\it ibid.} {\bf 8} 223; Cunningham E. 1909 {\it ibid.} {\bf 8} 77

\bibitem{FRW}  Fulton T., Rohrlich F. and Witten L. 1962 {\it Rev. Mod. Phys.}
{\bf 34} 442; 1962 {\it Nuovo Cimento} {\bf 26} 653

\bibitem{QSO95}  Jaekel M. T. and Reynaud S. 1995 {\it Quantum Semiclass.
Opt.} {\bf 7} 499

\bibitem{BJP95}  Jaekel M. T. and Reynaud S. 1995 {\it Brazilian J. Phys.}
{\bf 25} 325

\bibitem{Norton}  Norton J. D. 1993 {\it Rep. Progr. Phys.} {\bf 56} 791

\bibitem{JRloc}  Jaekel M. T. and Reynaud S. 1996 {\it Phys. Rev. Lett.}
{\bf 76} 2407; {\it Phys. Lett.} {\bf A220} 10; 1997 {\it EuroPhys. Lett.}
{\bf 38} 1; 1998 {\it Foundations of Physics} to appear
\end{references}
\end{document}